# Observation of lobes near the X-point in resonant magnetic perturbation experiments on MAST

A. Kirk, J. Harrison, Yueqiang Liu, E. Nardon[1], I.T. Chapman, P. Denner[2] and the MAST team

*EURATOM/CCFE Fusion Association, Culham Science Centre, Abingdon, Oxon OX14 3DB, UK.*
[1]*Association Euratom/CEA, CEA Cadarache, F-13108, St. Paul-lez-Durance, France*
[2]*University of York, Heslington, York YO10 5DD UK*

**Abstract**

The application of non-axisymmetric resonant magnetic perturbations (RMPs) with a toroidal mode number n=6 in the MAST tokamak produces a significant reduction in plasma energy loss associated with type-I Edge Localized Modes (ELMs), the first such observation with n>3. During the ELM mitigated stage clear lobe structures are observed in visible-light imaging of the X-point region. These lobes or manifold structures, that were predicted previously, have been observed for the first time in a range of discharges and their appearance is correlated with the effect of RMPs on the plasma i.e. they only appear above a threshold when a density pump out is observed or when the ELM frequency is increased. They appear to be correlated with the RMPs penetrating the plasma and may be important in explaining why the ELM frequency increases. The number and location of the structures observed can be well described using vacuum modelling. Differences in radial extent and poloidal width from vacuum modelling are likely to be due to a combination of transport effects and plasma screening.

The Edge-Localised Mode (ELM) is a repetitive instability associated with a steep pressure gradient, which can form at the edge of a tokamak plasma in high confinement regimes [1][2]. ELMs are explosive events, which can eject large amounts of energy and particles from the confined region [2]. In order to avoid damage to in-vessel components in future devices, such as ITER, a mechanism to suppress ELMs or reduce their size is required [3]. One such amelioration mechanism relies on perturbing the magnetic field in the edge plasma region, enhancing the transport of particles or energy and keeping the edge pressure gradient below the critical value that would trigger an ELM, while still maintaining an edge transport barrier. This technique of Resonant Magnetic Perturbations (RMPs) has been employed on DIII-D, where complete ELM suppression has been possible [4][5] and on JET [6][7], AUG [8] and MAST [9] where ELM mitigation (i.e. an increase in ELM frequency and decrease in ELM energy loss) has been obtained.

Although the ELM suppression observed in the low collisionality ITER-similar shaped plasmas on DIII-D can be explained by the reduction in the pedestal density and hence pressure gradient below that required to trigger an ELM [10], the increase in ELM frequency that is observed on JET and AUG and before complete suppression on DIII-D is more difficult to explain. In this Letter, the first experimental observation is reported of lobe-like structures, that extend into the Scrape Off Layer (SOL) near to the plasma X-point, in type I ELM-ing discharges on MAST. These plasma deformations may be important in explaining the ELM mitigation which is observed in these plasmas.

MAST is equipped with a set of in-vessel ELM control coils consisting of six coils above the mid-plane and twelve coils below [11]. Different coil configurations allow toroidal mode numbers in the range n=1 to 6 to be applied. Due to the up-down symmetry in the divertor coils on MAST, Single Null Diverted (SND) discharges are usually produced by shifting the plasma downwards. In this lower SND magnetic configuration the plasma is far from the upper row of RMP coils and hence the perturbation is predominantly from the lower row of coils, which produces a much broader spectrum of magnetic perturbation. If the RMPs are applied in an n=6 configuration with a current in the coils $I_{ELM}$ = 5.6 kAt, a clear increase in the ELM frequency and decrease of the ELM size is observed (see Figure 1). For $I_{ELM}$ < 3.2 kAt no effect is observed on the plasma, but above this threshold value the increase in ELM frequency is effectively linear. For the example shown in Figure 1, which has $I_{ELM}$ = 5.6 kAt, the ELM frequency increases from 80 Hz (for $I_{ELM}$ = 0 kAt) to 270 Hz while the energy lost from the core per ELM decreases from 16 kJ to 5 kJ. This is the first time that ELM mitigation has been observed on any device using a toroidal mode number greater than 3.

The radial pedestal profiles, obtained using Thomson scattering, show a drop in the pedestal density but little change in the electron temperature. The pedestal top pressure reduces but the peak pressure gradient remains unchanged. A stability analysis performed using the ELITE stability code [12] predicts that such a discharge would be stable to peeling-ballooning modes. Such a stability analysis assumes toroidally symmetric and smooth edge flux surfaces. However, as will be shown, during the application of RMPs the edge is anything but smooth and maybe it is these deformations of the surface that lead to greater instability.

The lower X-point region, shown in Figure 2a has been viewed using a toroidally viewing camera with a spatial resolution of 1.8mm at tangency plane. The image has been filtered with either an He II (468 nm) or a CIII (465 nm) filter and the images obtained using an integration time of 3000 or 300 µs respectively. These lines have been chosen since they are the strongest impurity lines in the typical plasma conditions found at the plasma boundary. Figure 2b shows a false colour image obtained using a He II filter at 0.32s in the shot with $I_{ELM}$ =0 kAt during an inter-ELM period. The image shows a smooth Last Closed Flux Surface (LCFS). In contrast, Figure 2c shows an image obtained at the same time during an inter-ELM period for the shot with $I_{ELM}$=5.6 kAt. Clear lobe structures are seen near to the X-point. Whilst these structures are clearest on the low field side (LFS) they are also visible on the high field side (HFS). These lobe structures only exist when the RMP coil current is above the threshold required to give an effect on the ELM frequency. For coil currents above the threshold ($I_{THR}$) the extent of the lobes increases approximately linearly with $I_{ELM}$-$I_{THR}$. These lobe structures have been observed in both L-mode and H-mode discharges with n=3, 4 or 6 but only when there is an accompanying effect of the RMPs on the plasma i.e. a density pump out in L-mode or an increase in ELM frequency in H-mode. Examples of these lobe structures in ELM mitigated H-mode discharges where the RMPs are applied in an n=4 or n=3 configuration are shown in Figure 3a and b respectively. While clear lobe structures are observed in each case, the location and poloidal separation of the lobes is different. For the same toroidal mode number of the perturbation the position of the lobes depends on the phase of the applied field (i.e. it depends on the sign of the current in the first coil).

The idea that so-called "Manifold" structures could exist was probably first introduced to the tokamak community by Evans et al., [13][14]. In an ideal axi-symmetric poloidally diverted tokamak the magnetic separatrix (or LCFS) separates the region of confined and open field lines. Non-axi-symmetric magnetic perturbations split this magnetic separatrix into a pair of so called "stable and unstable manifolds" [15]. Structures are formed where the manifolds intersect and these are particularly complex near to the X-point. The manifolds form lobes that are stretched radially both outwards and inwards. Some of these lobes can intersect the divertor target and result in the strike point splitting often observed during RMP experiments [16][17]. Although the lobes and strike point splitting require some RMP penetration, the observations of the strike point are typically made at a single toroidal angle and contain much less information on the magnetic field than the lobe images. In particular, it has been shown from modelling that the location of the structures observed at the strike point is not necessarily modified by the presence of significant RMP screening [18]. In contrast, RMP screening is expected to reduce the radial extent of the lobes (i.e. how far they extend from the LCFS), and therefore lobe images provide quantitative information to test plasma response models. In reference [19] it is shown that the radial extent of the lobes sets a minimum value on the radial extent of the stochastic layer, i.e. the stochastic layer has to be at least as broad as the lobes.

Calculations of what these lobes should look like have been performed based on numerical field line tracing using the ERGOS code [20] with external magnetic perturbations (from the in-vessel coils, intrinsic error field and ex-vessel error field correction coils) superimposed on the equilibrium plasma. Field lines are traced in both directions until they hit the divertor plate or complete 200 toroidal turns. A plot is then made in the poloidal plane of the location of each field line at a given toroidal location. The number of turns that they perform before hitting the plate is denoted by the colour of the dot. The pattern is independent of where the field lines started but is dependent on the toroidal location used for the plot. Figure 4a, b and c shows the resulting laminar plot at a toroidal angle ($\phi$) of 315 degrees for simulations with coil configurations producing perturbations with n=6, 4 and 3 respectively. The toroidal angle is chosen as the average toroidal location where the camera viewing chords are tangent to the unperturbed magnetic flux surface. The toroidal angle representing the tangency location varies from 305 degrees for pixels viewing the HFS of the plasma, through 315 degrees for pixels close to the X-point and up to 350 degrees for pixels at the far right hand side of the image.

A good quantitative agreement is observed between the number and separation of the lobes in the image and the modelling. In order to improve the comparison the laminar plots have been calculated in steps of 1 degree for the values of $\phi$ in the range 300 to 360 degrees for the case where the RMPs are in the n=6 configuration. For each pixel in the image the simulation at the nearest toroidal angle and R and Z location to the tangency point is chosen. A curve representing the boundary of the lobe structures produced is then superimposed on the image shown in Figure 5a. As can be seen the number and location of the lobes, particularly at the LFS, are in good agreement between the image and the modelling. At the HFS the agreement is less good and is possibly due to the effects of viewing through a greater volume of plasma and increased sensitivity to camera misalignment. There appears to be a discrepancy in the poloidal width of the lobes and their radial extent. This could be due to several effects: 1) smearing due to impurity light emission from other than the tangency location, 2) cross field diffusive transport and 3) plasma screening of the applied fields. Previous simulations using the EMC3 code [21] have shown that the strike point patterns can be broadened by cross field diffusion.

The effects of plasma screening have been approximated by reducing the applied value of $I_{ELM}$ used in the vacuum field simulations until the radial extent of the filaments in the simulations match those observed in the images. A value of $I_{ELM}$= 1.4 kAt, i.e. one quarter of the actual current, is found to be a good approximation as appears from the comparison shown in Figure 5b. A good agreement is observed in terms of the position and extent of the lobes. According to modelling [22][18], the observed large reduction of the lobes is possible only if resonant screening current sheets (RSCS) exist up to very close to the separatrix.

As mentioned above, the lobes are only observed for coil currents above a threshold $I_{THR}$. Furthermore, for $I_{ELM}>I_{THR}$ the radial extent of the lobes increases approximately linearly with $I_{ELM}-I_{THR}$. We speculate that the growth of the lobes with $I_{ELM}$ is due to the progressive disappearance of RSCS from the most external resonant surfaces towards the core as RMP penetration [23] occurs on these surfaces. However, the manifolds are expected to remain strongly reduced as long as there exists RSCS close enough to the separatrix, because these RSCS are very strongly coupled to the manifolds [18]. Hence, the manifolds are expected to appear only when $I_{ELM}$ is sufficient to remove all RSCS up to a large enough distance from the separatrix, which may be the cause of the observed threshold.

The effects of plasma screening on the laminar plots will be investigated in the future using the MARS-F code [24]. Previous MARS-F simulations of the effect of RMPs on the MAST plasma showed a clear correlation between the location of the maximum of the amplitude of the normal component of the plasma displacement at the plasma surface and the effect of the RMPs on the plasma [25]. Superimposed on the laminar plot in Figure 4a and c is the MHD plasma displacement calculated by MARS-F in the case of the n=6 and n=3 perturbations. The displacement from MARS-F has been scaled by a factor of 4 to match the overestimate found in the comparison of the images with the laminar plots (Figure 5b). MARS-F computes the linear response of a resistive plasma to the RMP field taking into account the experimentally measured toroidal flow to determine the screening. While the extent of the displacement is similar, the structure is smoother in the MARS-F simulations. This may partly be due to the fact that the MARS-F simulations, which are based on a flux co-ordinate system, have to smooth the X-point region and are only performed to a finite value of $m$ and hence exclude the very edge RSCS. It is not clear how the MHD displacement is related to the lobe structures observed and this will be the subject of further studies.

In conclusion, ELM mitigation MAST has been observed on MAST using RMPs with a toroidal mode number n=6. Coincident with the effect on the ELMs, for the first time, clear lobe-like structures are observed near to the X-point. The structures have been observed in MAST in both L and H-mode plasmas when resonant magnetic perturbations are applied. The appearance of these lobes is correlated with the effect of the RMPs on the plasma i.e. they only appear when the RMPs have an effect on the density in L-mode or on the ELM frequency in H-mode. The structures are not seen if the ELM coil current is too small or if the applied field is non-resonant. The number and location of the structures observed can be well described using vacuum modelling. However the radial extent is smaller in the images and the poloidal width larger. This is likely to be due to a combination of transport effects and plasma screening. Such 3D perturbations to the separatrix are not included in present stability codes. The inclusion of such effects in future codes may help to explain why the ELM frequency increases or why ELMs are suppressed when RMPs are applied.

## Acknowledgement

We would like to thank Pavel Cahyna of IPP Prague and Todd Evans of General Atomics for useful discussions on manifolds. This work, part-funded by the European Communities under the contract of Association between EURATOM and CCFE, was carried out within the framework of the European Fusion Development Agreement. The views and opinions expressed herein do not necessarily reflect those of the European Commission. This work was also part-funded by the RCUK Energy Programme under grant EP/I501045.

**Figures**

Figure 1 Time traces of a) the coil current ($I_{ELM}$), b) the line average density ($\bar{n}_e$) and the divertor $D_\alpha$ intensity for lower single null shots c) without and d) with RMP in an n=6 configuration.

Figure 2 a) Poloidal cross section of the MAST plasma with the view of the X-point camera marked as a rectangle. Images captured by the camera during an Inter-ELM period of a H-mode b) without RMPs and c) with RMPs in an n=6 configuration with $I_{ELM}$ = 5.6 kAt.

Figure 3 Images captured by the camera during an Inter-ELM period of a H-mode with RMPs in a) an n=4 and b) an n=3 configuration with $I_{ELM}$ = 5.6 kAt.

Figure 4 Laminar plots from ERGOS showing the number of toroidal turns performed before a field line intersects with the divertor for RMPs in a) n=6, b) n=4 and c) n=3 configurations. Superimposed on a) and c) as the black curve are the MARS-F calculations of the MHD plasma displacement scaled by a factor of 4.

Figure 5 Plasma image obtained using the RMPs in an n=6 configuration. Superimposed as a curve is the surface the lobe structures from the ERGOS simulations mapped on to the image using a) $I_{ELM}$ = 5.6 kAt and b) 1.4 kAt.

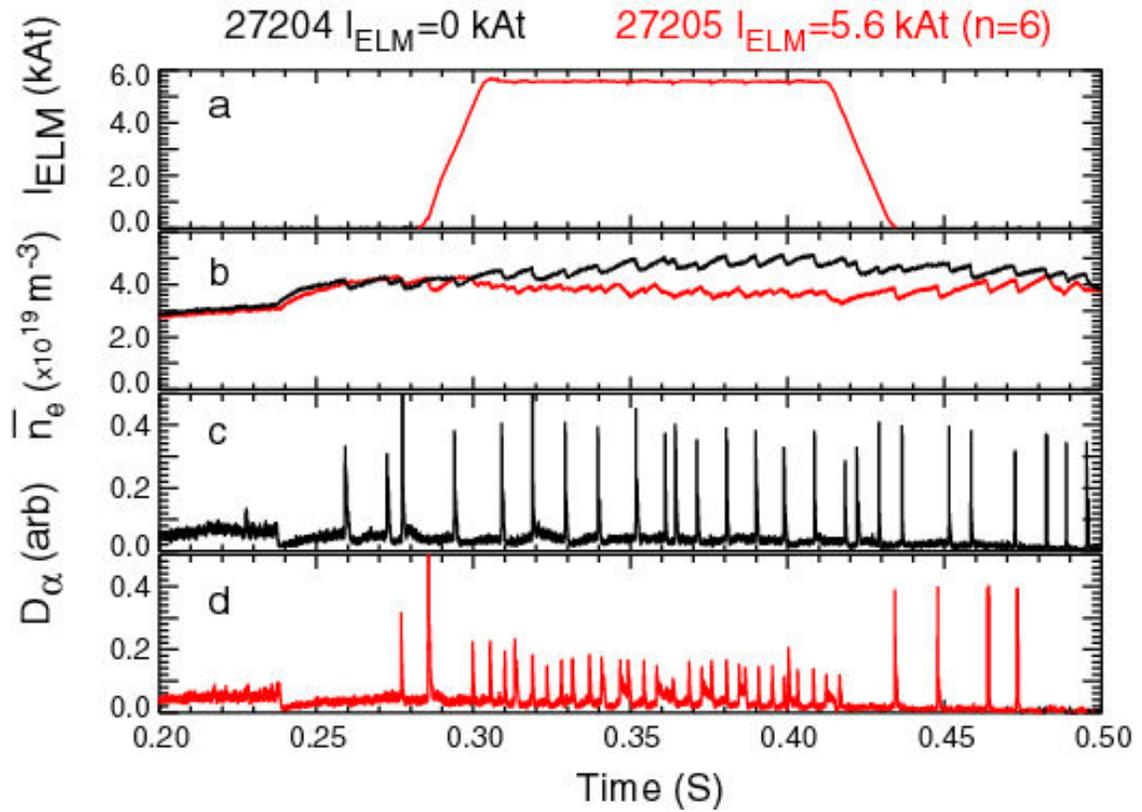

**Figure 1** Time traces of a) the coil current ($I_{ELM}$), b) the line average density ($\bar{n}_e$) and the divertor $D_\alpha$ intensity for lower single null shots c) without and d) with RMP in an n=6 configuration.

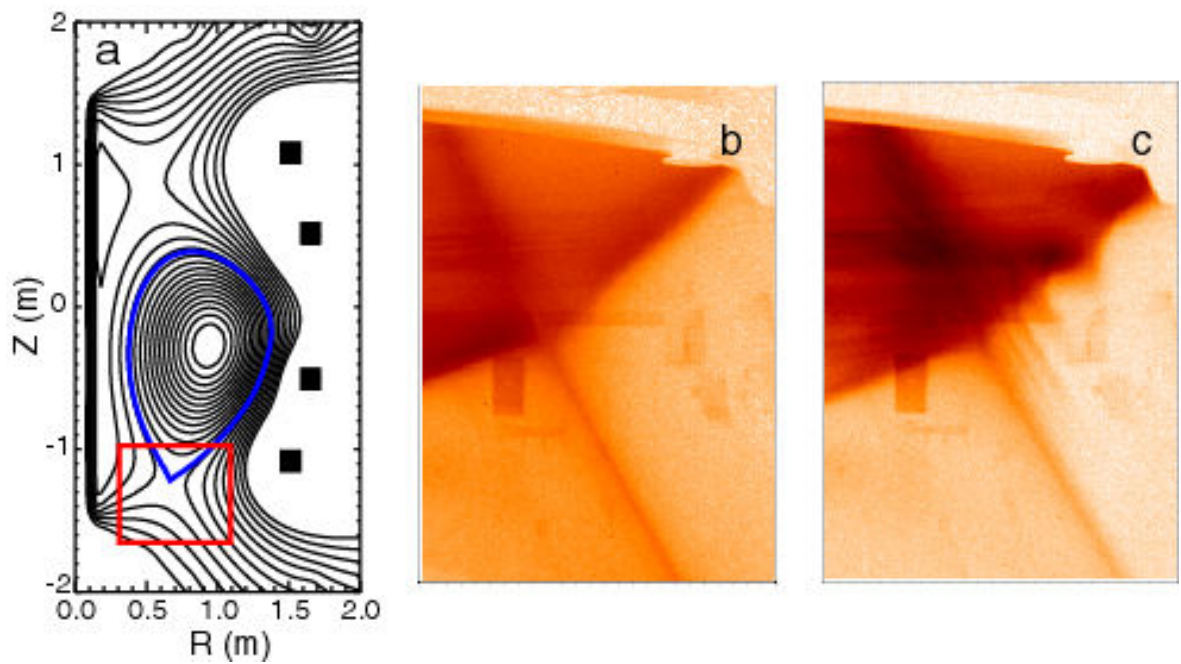

**Figure 2** a) Poloidal cross section of the MAST plasma with the view of the X-point camera marked as a rectangle. False colour images captured by the camera during an Inter-ELM period of a H-mode b) without RMPs and c) with RMPs in an n=6 configuration with $I_{ELM}$ = 5.6 kAt.

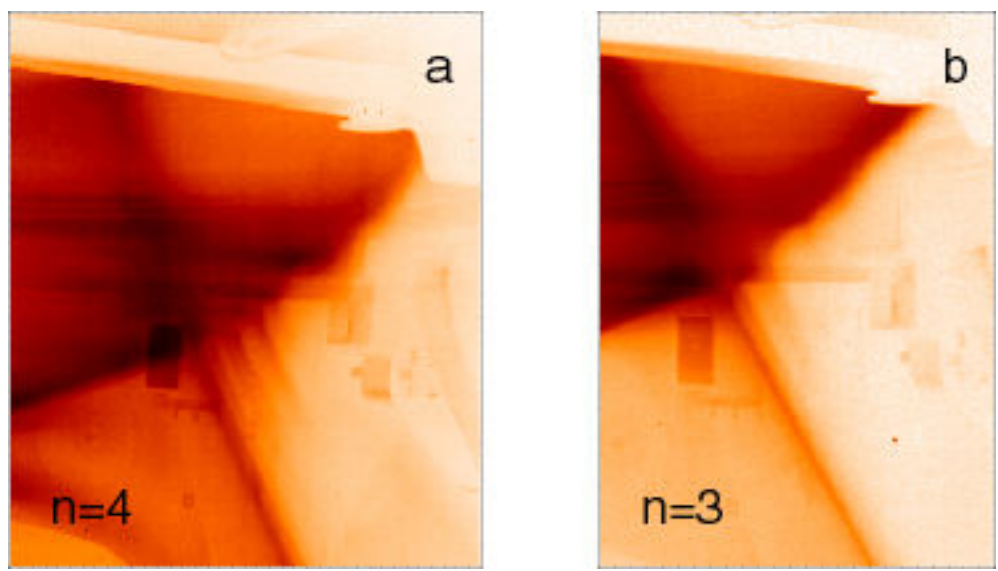

**Figure 3** False colour images captured by the camera during an Inter-ELM period of a H-mode with RMPs in a) an n=4 and b) an n=3 configuration with $I_{ELM}$ = 5.6 kAt.

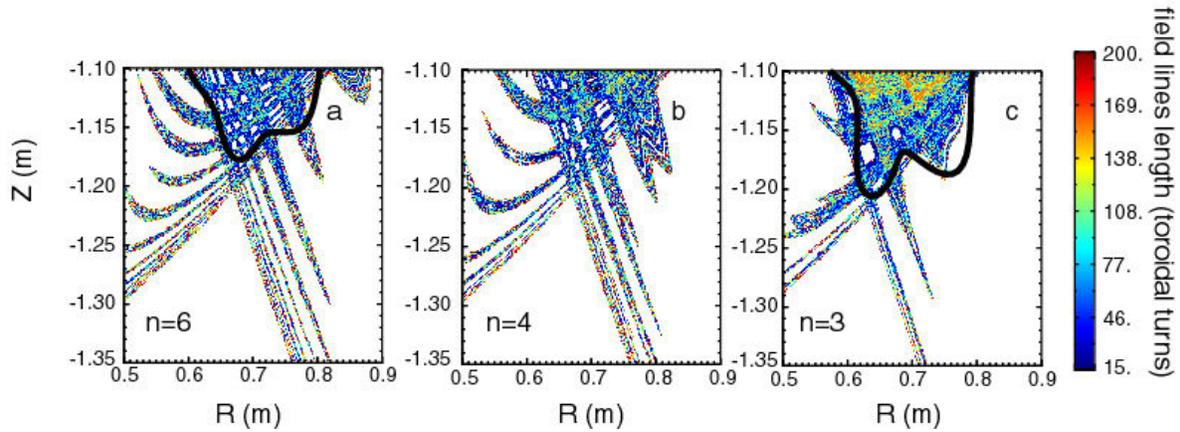

**Figure 4** Laminar plots from ERGOS showing the number of toroidal turns performed before a field line intersects with the divertor for RMPs in a) n=6, b) n=4 and c) n=3 configurations. Superimposed on a) and c) as the black curve are the MARS-F calculations of the MHD plasma displacement scaled by a factor of 4.

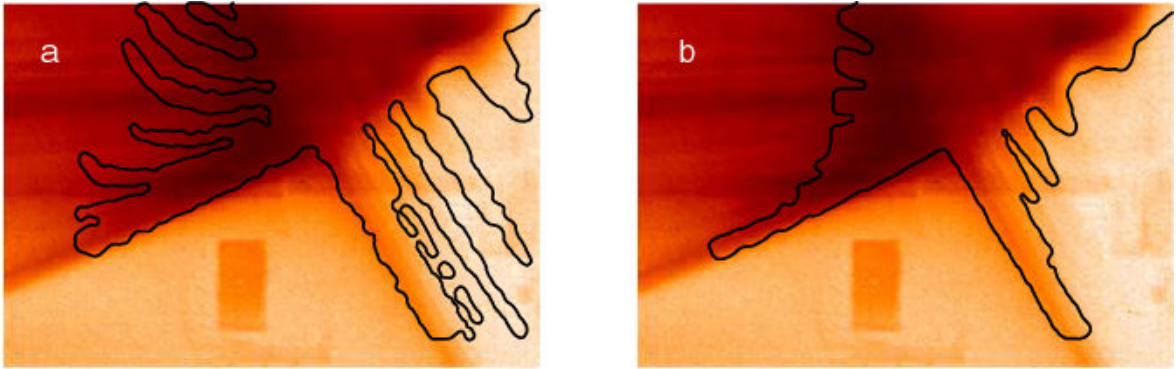

**Figure 5** Plasma image obtained using the RMPs in an n=6 configuration. Superimposed as a curve is the surface the lobe structures from the ERGOS simulations mapped on to the image using a) $I_{ELM}$ = 5.6 kAt and b) 1.4 kAt.